\documentclass[aps,prb,twocolumn,showpacs,superscriptaddress,floatfix]{revtex4-1}
\usepackage{graphicx}
\usepackage{amssymb}

\begin{document}

\title{Superconductivity in non-centrosymmetric YPtBi under pressure}

\author{T. V. Bay} \affiliation{Van der Waals - Zeeman Institute, University of Amsterdam, Science Park 904, 1098 XH Amsterdam, The Netherlands}
\author{T. Naka} \affiliation{National Institute for Materials Science, Sengen 1-2-1, Tsukuba, Ibaraki 305-0047, Japan}
\author{Y. K. Huang} \affiliation{Van der Waals - Zeeman Institute, University of Amsterdam, Science Park 904, 1098 XH Amsterdam, The Netherlands}
\author{A. de Visser} \email{a.devisser@uva.nl} \affiliation{Van der Waals - Zeeman Institute, University of Amsterdam, Science Park 904, 1098 XH Amsterdam, The Netherlands}

\date{\today}

\begin{abstract} We report a high-pressure single-crystal study of the non-centrosymmetric superconductor YPtBi ($T_c = 0.77$~K). Magnetotransport measurements show a weak metallic behavior with a carrier concentration $n \simeq 2.2 \times 10^{19}$~cm$^{-3}$. Resistivity measurements up to $p = 2.51$~GPa reveal superconductivity is promoted by pressure. The reduced upper critical field $B_{c2}(T)$ curves collapse onto a single curve, with values that exceed the model values for spin-singlet superconductivity. The $B_{c2}$ data point to an odd-parity component in the superconducting order parameter, in accordance with predictions for non-centrosymmetric superconductors.
\end{abstract}

\pacs{74.70.Dd, 74.62.Fj, 74.25.Op}

\maketitle

Recently, superconductivity with a transition temperature $T_c = 0.77$~K was discovered in the non-centrosymmetric half Heusler compound YPtBi~\cite{Butch2011}. Non-centrosymmetric (NCS) superconductivity (SC) forms a prominent research topic as it offers a wide-ranging, fruitful playground for the investigation of unconventional SC phases~\cite{*[{For a review see: }] [{ }] Bauer&Sigrist2012}. The lack of an inversion center in the crystal structure causes an electric field gradient, which creates an antisymmetric Rashba-type spin-orbit coupling. This results in a splitting of the Fermi surface, which thwarts spin-singlet or spin-triplet Cooper pairing of the conventional type. Instead, new pairing states, notably mixed even and odd parity Cooper pair states, are predicted to make up the SC condensate~\cite{Frigeri2004}. The field of NCS SC was initiated by the discovery of SC in the heavy-fermion material CePt$_3$Si ($T_c = 0.75$~K)~\cite{Bauer2004}. Other well-documented examples of NCS SCs are: CeRhSi$_3$ ($T_c = 1.1$~K under pressure)~\cite{Kimura2005}, CeIrSi$_3$ ($T_c = 1.6$~K under pressure)~\cite{Sugitani2006}, Li$_2$Pt$_3$B ($T_c = 2.6$~K)~\cite{Togano2004} and Mo$_3$Al$_2$C ($T_c = 9.2$~K)~\cite{Bauer2010,Karki2010}. For several NCS SCs solid evidence for an odd-parity component in the SC order parameter has been extracted from critical magnetic fields exceeding the spin-singlet Pauli paramagnetic limit and/or line or point nodes in the SC gap function~\cite{*[{For a review see: }] [{ }] Bauer&Sigrist2012}.

Yet another motive to investigate YPtBi is provided by electronic band structure calculations~\cite{Chadov2010,Lin2010}. First principle calculations carried out on a series of rare earth (RE) ternary half Heusler compounds predicted several of them to have a topologically non-trivial band structure, due to a sizeable $\Gamma _6 - \Gamma _8$ band inversion. This allows for a classification as (candidate) 3D topological insulator. A 3D topological insulator is insulating in the bulk, but has conducting surface states protected by a non-trivial $Z _2$ topology~\cite{Hasan&Kane2010,Qi&Zhang2010}. Notably, the non-magnetic REPtBi compounds LuPtBi and LaPtBi, as well as YPtBi, are predicted to have a relatively strong band inversion. Transport experiments reveal LuPtBi is metallic~\cite{Mun2010}, while YPtBi~\cite{Canfield1991,Butch2011} and LaPtBi~\cite{Goll2008} are semimetals that become SC ($T_c = 0.9$~K for LaPtBi~\cite{Goll2008}). The non-trivial topology of the electron bands makes YPtBi and LaPtBi promising candidates for topological SC with protected Majorana surface states~\cite{Schnyder2012}. Topological SCs are rare, and only a few cases are known: the correlated metal Sr$_2$RuO$_4$~\cite{MacKenzie&Meano2003}, which is a time-reversal symmetry-breaking chiral 2D $p$-wave SC~\cite{Kitaev2009}, and the intercalated thermoelectric effect material Cu$_x$Bi$_2$Se$_3$~\cite{Hor2010,Kriener2011a}, for which a time reversal symmetric, fully-gapped, odd-parity superconducting state has been proposed~\cite{Fu&Berg2010}.

YPtBi has a cubic structure and crystallizes in the $F\overline{4}3m$ space group. It was first prepared as non-$f$ electron reference material in the systematic investigation of magnetism and heavy-fermion behavior in the REPtBi series~\cite{Canfield1991}. Magnetotransport measurements carried out on single crystals grown out of Bi flux point to semimetallic-like behavior~\cite{Canfield1991,Butch2011}. The resistivity, $\rho (T)$, increases steadily upon cooling below 300~K and levels off below $\sim$ 60~K. The  Hall coefficient $R_H$ is positive and quasi-linear in the magnetic field, allowing for an interpretation in a single-band model with hole carriers. The carrier concentration, $n_h$, is low and shows a substantial decrease upon cooling from $ 2 \times 10^{19}$~cm$^{-3}$ at 300~K to $ 2 \times 10^{18}$~cm$^{-3}$ at 2~K. Concurrently, the Sommerfeld coefficient in the specific heat is very small, $\gamma \leq 0.1$~mJ/molK$^2$ (Ref.~\onlinecite{Pagliuso1999}). YPtBi is diamagnetic and the magnetic susceptibility, $\chi$, attains a temperature independent value of $-10^{-4}$ emu/mol (Ref.~\onlinecite{Pagliuso1999}).

The transition to the SC state takes place at $T_c = 0.77$~K~\cite{Butch2011}, where the resistivity sharply drops to 0. At the same temperature a diamagnetic screening signal appears in the  ac-susceptibility, $\chi _{ac}$, but the magnetic response is sluggish. The upper critical field, $B_{c2} (T)$, shows an unusual quasi-linear behavior and attains a value of $\sim$1.5~T for $T \rightarrow 0$~K~\cite{Butch2011}. Heat capacity measurements around the normal-to-SC phase transition, which are a standard experimental tool to provide evidence for bulk SC, have not been reported yet. Notice, the extremely small $\gamma$-value makes this a difficult experiment. On the other hand, the confirmation of SC with a critical temperature $T_c = 0.77$~K in our single crystals and the observation of a diamagnetic signal in $\chi _{ac}$ which corresponds to a SC volume fraction of 30\%~\footnote{T. V. Bay, unpublished.}, point to a bulk SC phase.

\begin{figure}

\includegraphics[width=8cm]{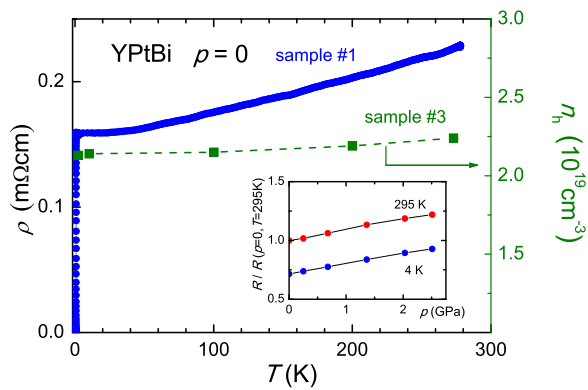}
\caption{(color online) Resistivity (closed circles) and carrier concentration (closed squares - right axis) of YPtBi as a function of temperature at ambient pressure. Inset: Resistance at 295~K and 4~K as a function of pressure. Resistance values are normalized to the room temperature value at ambient pressure, $R(p=0,T=295$~K).}
\end{figure}

Here we report the response of the SC phase of YPtBi to pressures up to 2.51~GPa. Transport measurements on single crystals confirm SC with a critical temperature $T_c = 0.77$~K. Under pressure SC is enhanced and $T_c$ increases at a linear rate of $0.044$~K/GPa. The upper critical field $B_{c2}(T)$ curves taken at different pressures collapse onto a single curve, with values that exceed the model values for spin-singlet SC. The $B_{c2}$ data point to the presence of an odd-parity Cooper pairing component in the SC order parameter, in agreement with predictions for NCS and topological SCs.~\cite{Frigeri2004,Schnyder2012,Fu&Berg2010}

Several batches of YBiPt were prepared out of Bi flux. Powder x-ray diffraction confirmed the $F\overline{4}3m$ space group. The lattice parameter $a= 6.650$~{\AA}~ in good agreement with literature~\cite{Butch2011}. Single crystals taken from these batches showed reproducibly SC with a resistive transition at $T_c = 0.77$~K. The resistivity and Hall effect were measured using a MaglabExa system (Oxford Instruments) for $T=$ 4 - 300~K and in a $^3$He refrigerator (Heliox, Oxford Instruments) for 0.24 - 10~K. Additional $\rho (T)$ data were taken in a dilution refrigerator (Kelvinox, Oxford Instruments) down to $T= 0.04$~K. The transport data were measured using a low-frequency ($f = 13$~Hz) lock-in technique with a low excitation current ($I = 100 ~\mu$A). The high-pressure transport measurements were carried out using a hybrid clamp cell made of NiCrAl and CuBe alloys. Samples were mounted on a plug which was placed in a Teflon cylinder with Daphne oil 7373 as hydrostatic pressure transmitting medium. The pressure cell was attached to the cold plate of the $^3$He refrigerator. The pressure was determined in situ by the SC transition of a Sn specimen.

\begin{figure}
\includegraphics[width=8cm]{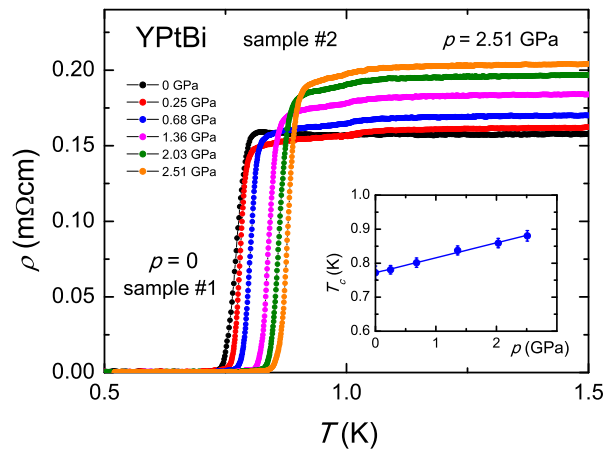}
\caption{(color online) Superconducting transition of YPtBi at pressures of 0, 0.25, 0.68, 1.36, 2.03 and 2.51~GPa (from left to right). Data at $p=0$ taken on sample \#1; data under pressure on sample \#2. Inset: Superconducting transition temperature as a function of pressure. The solid line is a linear fit to the data points with slope $dT_c /dP = 0.044$~K/GPa.}
\end{figure}

In Fig.~1 we show a typical resistivity trace $\rho (T)$. Upon cooling below 300 K $\rho (T)$ gradually drops and levels off below 30~K. This demonstrates our YPtBi crystals behave as a metal, rather than as a semimetal~\cite{Canfield1991,Butch2011}. The carrier concentration $n_h (T)$ is low and displays a weak temperature variation (Fig.~1). Near room temperature the transport parameters of our samples are quite similar to those reported in Ref.~\onlinecite{Butch2011}: $\rho (300$~K) equals 230~$\mu \Omega$cm \textit{versus} 300~$\mu \Omega$cm and $n_h (300$~K) equals $ 2.2 \times 10^{19}$~cm$^{-3}$ \textit{versus} $ 2 \times 10^{19}$~cm$^{-3}$. A major difference is found in $n_h (T)$, which is close to temperature independent for our sample, but drops a factor 10 upon cooling to 2~K in Ref.~\onlinecite{Butch2011}. The origin of the dissimilar transport behavior is unclear. Possibly trapping of carriers at defects upon lowering the temperature causes semimetallic-like behavior in some of the samples. The metallic behavior is robust to pressure (see the inset in Fig.~1). $R(300$~K) increases linearly with pressure, resulting in a 20\% increase at the maximum pressure of 2.51~GPa. The residual resistance $R(4$~K) increases at the same rate. The residual resistance ratio, $RRR= R(300$~K)/$R(4$~K), of our samples amounts to $\sim$ 1.4 at $p=0$. A sharp SC transition is observed for all samples at $T_c = 0.77$~K. The width of the transition $\Delta T_c $, as determined between 10\% and 90\% of the normal state $R$-value, is 0.06~K.

The SC transition under pressure is shown in Fig.~2. Notice, the $p=0$ data are taken on a different sample in a separate experiment. $T_c$, as determined by the maximum in the slope $d \rho / dT$ increases linearly with pressure at a rate of 0.044~K/GPa (see inset in Fig.~2). The width of the transition does not change with pressure, which is indicative of a homogeneously applied pressure. The $\rho (T)$ data taken on sample \#2 (under pressure) show a tiny structure just above 1~K. This feature is insensitive to pressure and suppressed by a small magnetic field ($B \sim 0.1$~T, see Fig.~3). It has not been observed in other samples.

The relatively weak pressure dependence of $\rho (T)$ and the enhancement of $T_c$ with pressure are unexpected for a low carrier density material. For instance in Cu$_x$Bi$_2$Se$_3$, which has a comparable metallic behavior and low carrier concentration, the resistance is enhanced and $T_c$ decreases under pressure~\cite{Bay2012}. In that case the variation $T_c (p)$ can be understood qualitatively in a simple model, where $T_c \sim \Theta_D \exp [-1/N(0)V_0]$, with $\Theta_D$ the Debye temperature, $N(0) \sim m^* n^{1/3}$ the density of states (with $m^*$ the effective mass) and $V_0$ the effective interaction parameter~\cite{Cohen1969}. For Cu$_x$Bi$_2$Se$_3$ $n$ decreases with pressure, and accordingly $T_c$ decreases~\cite{Bay2012}. For YPtBi, the weak variation of $R$ with pressure (Fig.~1), suggests $n$ is close to pressure independent.  Therefore, the increase $T_c (p)$ indicates the product $N(0)V_0$ has a more involved dependence on pressure.

\begin{figure}
\includegraphics[width=8cm]{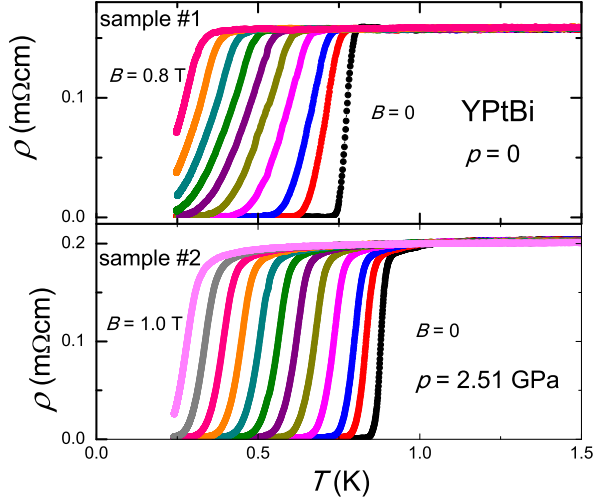}
\caption{(color online) Superconducting transition of YPtBi measured in magnetic fields of 0, 0.05, 0.1, 0.2, 0.3, 0.4, 0.5, 0.6, 0.7, 0.8, 0.9 and 1.0 T (from right to left). Upper frame: $p=0$,  sample \#1. Lower frame: $p=2.51$~GPa, sample \#2.}
\end{figure}

The depression of SC by a magnetic field was measured in fixed fields up to $p=$~2.51~GPa. Representative data are shown in Fig.~3. For sample \#1 measured at $p=0$ $\Delta T_c$ increases almost a factor 2 to 0.12~K in the highest field. For sample \#2, measured under pressure, $\Delta T_c$ is virtually pressure and field independent, which attests its high quality. $T_c (B)$ determined by the maximum in $d \rho /dT$ at fixed $B$ is reported for each pressure in Fig.~4. $B_{c2} (T)$ is dominated by a quasi-linear temperature dependence down to $T_c /3$. At temperatures there below, $B_{c2} (T)$ curves towards the vertical axis. For $p=0$ we obtain $B_{c2} (T \rightarrow 0) \simeq 1.23$~T. Notice, close to $T_c$ all data sets show a weak curvature or tail. The curvature is less pronounced for the better sample (\#2) measured under pressure.

Next we extract parameters that characterize the SC state and investigate whether our samples are sufficiently pure to allow for odd-parity superconductivity~\cite{Balian&Werthamer1963}. From the relation $B_{c2} = \Phi _0 / 2 \pi \xi ^2$, where $\Phi _0$ is the flux quantum, we calculate a SC coherence length $\xi = 17$~nm. An estimate for the electron mean free path, $\ell$, can be obtained from the relation $\ell = \hbar k_F /\rho _0 n e^2$, assuming a spherical Fermi surface $S_F = 4 \pi k_F ^2$ with Fermi wave number $k_F =(3 \pi ^2 n)^{1/3}$. With $n = 2.2 \times 10^{25}$~m$^{-3}$ and $\rho _0 = 1.6 \times 10^{-6}~ \Omega$m (see Fig.~1) we calculate $k_F = 0.9 \times 10^9$~m$^{-1}$ and $\ell = 105$~nm. Thus $\ell > \xi$, which tells us YPtBi is in the clean limit. Similar values for $\ell$ and $\xi$ were obtained in Ref.~\onlinecite{Butch2011}. A more elaborated analysis can be made by employing the slope of the upper critical field $dB_{c2}/dT$ at $T_c$~\cite{Orlando1979}: $|dB_{c2}/dT|_{T_c} \simeq ~4480 \cdot \gamma \rho_0 + ~1.38 \times 10^{35} \cdot \gamma ^2 T_c / S_F ^2$. Assuming $\gamma \sim n^{1/2}$ we estimate for our sample $\gamma =~ $7.3~J/m$^3$K$^2$ based on the value of 2.3~J/m$^3$K$^2$ (Ref.~\onlinecite{Pagliuso1999}) and by taking into account that for our sample $n$ at low $T$ is $10 \times$ higher than reported in Ref.~\onlinecite{Butch2011}. With the experimental values $\rho _0 = 1.6 \times 10^{-6}~ \Omega$m, $T_c = 0.77$~K and $|dB_{c2}/dT|_{T_c} = 1.9$~T/K (see Fig.~4, we neglect the weak curvature close to $T_c$) we calculate $k_F = 0.4 \times 10^9$~m$^{-1}$, $\xi = 20$~nm and $\ell = 582$~nm. This confirms $\ell > \xi$. The weak pressure response of the transport parameters justifies the conclusion that the clean limit behavior is also obeyed under pressure.

\begin{figure}
\includegraphics[width=8cm]{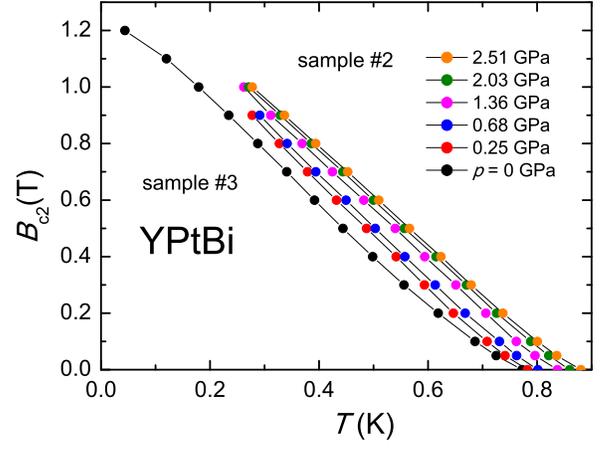}
\caption{(color online) Temperature variation of the upper critical field $B_{c2}(T)$ at pressures of 0, 0.25, 0.68, 1.36, 2.03 and 2.51 GPa (from bottom to top). Data at $p=0$ are taken on sample \#3; data under pressure on sample \#2. }
\end{figure}

For a standard weak-coupling spin-singlet SC in the clean limit the orbital critical field is given by $B_{c2}^{orb}(0)=0.72 \times T_c ~|dB_{c2}/dT|_{T_{c}}$ (Werthamer-Helfand-Hohenberg [WHH] model~\cite{Werthamer1966}). If one considers in addition the suppression of the spin-singlet state by paramagnetic limitation~\cite{Clogston1962,Chandrasekhar1962}, the resulting critical field is reduced to $B_{c2}(0)=B_{c2}^{orb}(0)/\sqrt{1 + \alpha ^2}$, with the Maki parameter $\alpha = \sqrt {2} B_{c2}^{orb}(0)/ B^P (0)$~\cite{Werthamer1966,Maki1966} and the Pauli limiting field $B^P (0) = 1.86 \times T_c$. For YPtBi we calculate $B_{c2}^{orb}(0)=1.05$~T, $B^P (0) = 1.43$~T, $\alpha = 1.04$ and $B_{c2}(0) = 0.73$~T. The latter value is much lower than the experimental value $B_{c2}(0) = 1.24$~T and we conclude $B_{c2}$ is dominated by the orbital limiting field.

\begin{figure}
\includegraphics[width=7cm]{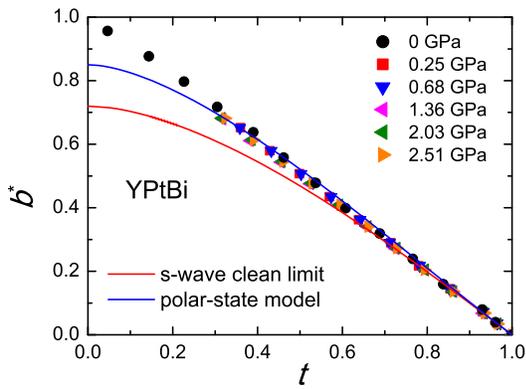}
\caption{(color online) Reduced upper critical field $b^*= (B_{c2}(T)/T_c)/|dB_{c2}/dT|_{T_c}$ as a function of the reduced temperature $t= T/T_c$ at pressures of 0, 0.25, 0.68, 1.36, 2.03 and 2.51 GPa. Notice, we neglected the small tail close to $T_c$ and obtained $|dB_{c2}/dT|_{T_c}$ from the field range $B = 0.1-0.2$~T. The red (lower) and blue (upper) full lines represent model calculations for an $s$ and $p$-wave superconductor (see text). }
\end{figure}

In Fig.~5 we present the $B_{c2}$ data at different pressures in a reduced plot $b^*(t)$, with $b^* = (B_{c2}/T_{c})/|dB_{c2}/dT|_{T_{c}}$ and $t = T/T_c$ the reduced temperature. All the $B_{c2}(T)$ curves collapse onto a single function $b^*(t)$. In Fig.~5 we have also traced the universal $B_{c2}$ curve for a clean orbital limited spin-singlet SC within the WHH model~\cite{Werthamer1966}. Clearly, the data deviate from the standard spin-singlet behavior. Notably the fact that our $B_{c2}$ data are well above even these universal values is a strong argument in favor of odd-parity SC. A similar conclusion based on $B_{c2}$ data was drawn for the candidate topological superconductor Cu$_x$Bi$_2$Se$_3$~\cite{Bay2012}. Finally, we compare the $B_{c2}(T)$ data with the polar-state model function of a spin-triplet SC~\cite{Scharnberg&Klemm1980}. Overall, the $B_{c2}$ values match the model function better, but significant discrepancies remain. Notably, the unusual quasi-linear $b^* (t)$ down to $t /3$ is not accounted for, while below $t/3$ the data exceed the model function values. Clearly, more theoretical work is needed to capture the intricate behavior of mixed spin-singlet and spin-triplet superconductors in an applied magnetic field.

In summary, we have prepared single crystals of the non-centrosymmetric superconductor YPtBi. Superconductivity is confirmed at $T_c = 0.77$~K. Transport measurements demonstrate our crystals exhibit metallic rather than semimetallic behavior. We have investigated the response to pressure  of the superconducting phase and find superconductivity is promoted by pressure. The upper-critical field $B_{c2}$ data under pressure collapse onto a single universal curve, which differs from the standard curve of a weak-coupling, orbital-limited, spin-singlet superconductor. The sufficiently large mean free path, the absence of Pauli limiting and the unusual temperature variation of $B_{c2}$, all point to a dominant odd-parity component in the superconducting order parameter of non-centrosymmetric YPtBi.

Acknowledgements $-$ T.V. Bay acknowledges support of the Vietnamese Ministry of Education and Training.

\bibliography{RefsTI}

\end{document}